# FERMI-HUBBARD PHYSICS WITH ATOMS IN AN OPTICAL LATTICE[1]

**Tilman Esslinger, Department of Physics, ETH Zurich, Switzerland**

**ABSTRACT**

The Fermi-Hubbard model is a key concept in condensed matter physics and provides crucial insights into electronic and magnetic properties of materials. Yet, the intricate nature of Fermi systems poses a barrier to answer important questions concerning d-wave superconductivity and quantum magnetism. Recently, it has become possible to experimentally realize the Fermi-Hubbard model using a fermionic quantum gas loaded into an optical lattice. In this atomic approach to the Fermi-Hubbard model the Hamiltonian is a direct result of the optical lattice potential created by interfering laser fields and short-ranged ultracold collisions. It provides a route to simulate the physics of the Hamiltonian and to address open questions and novel challenges of the underlying many-body system. This review gives an overview of the current efforts in understanding and realizing experiments with fermionic atoms in optical lattices and discusses key experiments in the metallic, band-insulating, superfluid and Mott-insulating regimes.

**INTRODUCTION**

The previous years have seen remarkable progress in the control and manipulation of quantum gases. This development may soon profoundly influence our understanding of quantum many-body physics. The quantum gas approach to many-body physics is fundamentally different from the path taken by other condensed matter systems, where experimentally observed phenomena trigger the search for a theoretical explanation. For instance, the observation of superconductivity resulted in the Bardeen-Cooper-Schrieffer (BCS) theory. In quantum gas research, the starting point is a many-body model that can be realized experimentally, such as a weakly interacting Bose gas. The study of the system then leads to the observation of a fundamental phenomenon like Bose-Einstein condensation (BEC) (1; 2; 3). Other examples are the crossover from a molecular Bose-Einstein condensate to the BCS regime (4; 5; 6; 7; 8; 9; 10), and the superfluid to Mott-insulator transition in a Bose gas exposed to an optical lattice potential (11; 12).

The outstanding challenge for the research field of quantum gases is to gain distinctive and new insights into quantum many-body physics. For example, will it be possible to answer long-standing questions of an underlying model? A unique system in the quest for answers is the Fermi-Hubbard model, which is a key model to describe electronic properties of solids. It

---

assumes a single static band and local interactions between the particles (13; 14; 15). Yet, the question as to whether the ground state of the two-dimensional Fermi-Hubbard model supports d-wave superfluidity or superconductivity (16) has so far defied theoretical explanation - despite 20 years of intensive and very fruitful efforts in condensed matter physics. Similarly, spin frustrated Hubbard models and Hubbard models with disorder carry open puzzles. These difficulties originate in the interplay of localization, coherence and spin ordering (17).

Therefore, increasing research efforts in the field of quantum gases are directed towards the study of a repulsively interacting two-component Fermi gas trapped in the periodic potential of a three-dimensional optical lattice. This system provides an almost ideal experimental realization of the Fermi-Hubbard model with highly tunable parameters. The expectation is that an experiment may provide a diagram of the low-temperature phases of the two-dimensional Fermi-Hubbard model (18). Such an achievement can be regarded as the quantum simulation of an open and highly relevant problem in quantum many-body physics. The research on quantum gases in optical lattices has to be viewed naturally in the context of the theoretical efforts. The concept of a Hubbard model realized through a quantum gas in an optical lattice was originally proposed in a theoretical publication (11). Its experimental realization for bosons (12), fermions (19) and Bose-Fermi mixtures (20; 21) has in turn triggered tremendous theoretical interest into these systems.

Experiments have now accessed the strongly correlated physics of the Fermi-Hubbard model (22; 23; 24). It therefore appears appropriate to give a perspective on the state of current efforts and to discuss the future prospects. A further goal of this review is to make basic concepts and considerations relevant to these experiments accessible to readers from the condensed matter community and researchers entering the field. The review focuses on fermionic quantum gases in optical lattices and their link to the Fermi-Hubbard model. Previous reviews on optical lattices have only briefly discussed the Fermi-Hubbard model. Atomic realizations of strongly interacting quantum many-body systems have been reviewed in references (25; 26), the optical lattice as a Hubbard tool box has been examined in reference (27), and the dynamics in optical lattices has been discussed in reference (28). General reviews on trapped bosonic and fermionic quantum degenerate gases can be found in references (3; 29; 30; 31).

After an introduction to the fundamentals of quantum gas research I will briefly present the general concept of an optical lattice. This will be followed by a discussion of the Fermi-Hubbard model, with emphasis on the consequences of the inhomogenous trapping potential. Subsequently, I will look at experimental realizations of Fermi gases in three dimensional optical lattices and recent progress in the strongly correlated regime. The major challenges for future progress are to reach substantially lower temperatures, to access correlation functions and to facilitate local readout. I discuss the progress towards these goals at the end of the review.

**Atomic Quantum Gases**

The route towards quantum degeneracy in a gas of atoms is a narrow path which circumvents solidification by keeping the atomic cloud at densities below $10^{14}$ atoms per cm$^3$ (32). At higher densities, inelastic three-body collisions set in, which lead to the formation of molecules and the release of the binding energy. To go along this track, the experiments exploit the mechanical effect of laser light to slow down atoms (33; 34; 35), followed by evaporative cooling of the atomic cloud in a magnetic or optical trap (32). These efforts are rewarded with an experimental quantum many-body system which can be fully characterized at the microscopic level and is decoupled from the environment. The first demonstrations and investigations of Bose-Einstein condensation in a weakly interacting gas of atoms were soon accompanied by the successful creation of quantum degenerate Fermi gases (36; 37; 38). Exploiting Feshbach resonances (39; 40) has made it possible to tune the collisional interaction between the atoms and enter the crossover regime between a molecular BEC and BCS pairing (29; 30). Quantum degenerate Fermi gases have so far been prepared in the following atomic species: $^{40}$K (36), $^6$Li (37; 38), $^3$He* (41) and $^{173}$Yb (42).

**Atoms in Optical Lattices**

An intriguing tool to manipulate ultracold quantum gases is the optical lattice (27) which is created by laser standing waves. In the optical lattice, atoms experience a periodic potential due to the interaction of the induced electric dipole moment of the atoms with the laser light (43). For frequencies of the lattice laser which are below the atomic resonance frequency, the atoms are attracted towards regions of high laser intensity. This is often referred to as red detuning. In this case, an anti-node of the standing wave acts as a potential minimum.

The concept of an optical lattice was originally proposed in the context of laser spectroscopy (44; 45), and with the invention of laser cooling (46; 47), it became experimentally accessible. The quantization of the atomic motion in optical lattice potentials was investigated (48; 49; 50; 51), and the periodic arrangement of the atoms was directly revealed by observing laser light scattered under the Bragg-angle from this very dilute atomic sample (52; 53). Furthermore, Bloch oscillations (54) and the Wannier Stark ladder (55) could be observed with laser cooled atoms in far detuned optical lattices.

By loading Bose-Einstein condensates into the periodic potential of a single laser standing wave, a new direction was explored. After initial studies on the coherence properties of the Bose-Einstein condensate (56; 57), a wide range of research was performed in this configuration (28). Furthermore, the research has been extended to quantum degenerate Fermi gases (58) and to Bose-Fermi mixtures (59).

The strongly correlated physics of the Bose-Hubbard model became accessible in an experimental set-up with three standing laser waves intersecting at the position of the condensate. The predicted quantum phase transition (11; 60) between a spatially coherent superfluid and a Mott-insulating state showed up in a sudden disappearance of the spatial coherence when the ratio between collisional and kinetic energy reached a critical value (12). The reversibility of this process and the gapped excitation spectrum of the Mott-insulator were subject of early studies (12; 61; 62). Another experimental path was to freeze the atomic motion along two directions and to study one-dimensional Bose gases (63; 64; 65; 66). Many fascinating aspects of these strongly correlated Bose systems have been reviewed in reference (25).

More recently, an atomic Fermi-Hubbard model could be realized by loading a quantum degenerate gas of fermionic atoms into a three-dimensional optical lattice potential (19). This has directly connected the research frontiers in quantum gas and condensed matter physics.

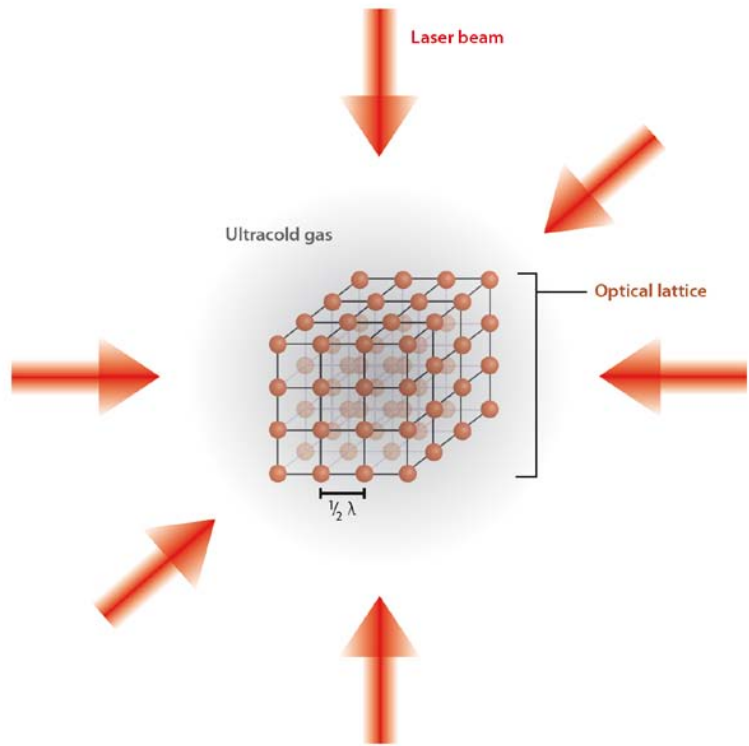

*Figure 1*: Three-dimensional optical lattice. An ultracold gas of atoms is trapped in the overlap region of three pairs of counter-propagating laser beams. Each pair produces a standing laser wave in which the atoms experience a periodic potential. All three pairs generate a three-dimensional simple cubic lattice structure, with the separation between adjacent lattice sites being half of the laser wavelength λ. In addition, the Gaussian beam profile gives rise to a force pointing towards the beam center, where the atoms are harmonically confined. The typical number of trapped atoms in current experiments is between $10^4$ and $10^6$. The periodicity of the optical lattice results in a band structure for the trapped atoms. The physics of an interacting quantum gas in the optical lattice can often be described by a Hubbard model.

**THE FERMI-HUBBARD MODEL IN AN ATOM TRAP**

An optical lattice created by three mutually perpendicular laser standing waves gives rise to a periodic potential of the form $V_{lat} = V_{0x}\cos^2(kx) + V_{0y}\cos^2(ky) + V_{0z}\cos^2(kz)$, where $k=\lambda/2\pi$ is the wave vector of the lattice laser, see figure 1. The lattice constant *d* is related to the laser wavelength $\lambda$ by $d=\lambda/2$. A symmetric lattice corresponds to $V_0 = V_{0x,y,z}$. The lattice depths $V_0$ is proportional to the laser intensity and inversely proportional to the detuning between laser frequency and atomic transition frequency (43). The values $V_{0x,y,z}$ are usually assumed to be constant, which is justified for a region in the trap center, which is small compared with the Gaussian beam waists. Interference terms between the three standing waves can be avoided by choosing suitable polarizations and frequency offsets for the standing waves (12). The lattice depth $V_0$ is often expressed in terms of the recoil energy $E_r = (\hbar k)^2/2m$, which is the kinetic energy of an atom with mass *m* and the momentum $\hbar k$ of a single lattice photon.

For deep enough lattice potentials, the atomic field operators can be expanded in terms of localized Wannier functions, and the atomic motion is determined by tunneling between adjacent sites. In the lowest band, direct tunneling to next to nearest neighbors is typically suppressed by one order of magnitude compared with nearest neighbor tunnelling (11). In most experiments, the Fermi gas is prepared in a mixture of two spin states, which correspond to different magnetic sublevels of the atomic ground state. The collision between two atoms in different spin states residing on the same lattice site gives rise to a short ranged interaction. With all atoms prepared in the lowest band, this concept leads to the Hubbard Hamiltonian (11) and is illustrated in figure 2.

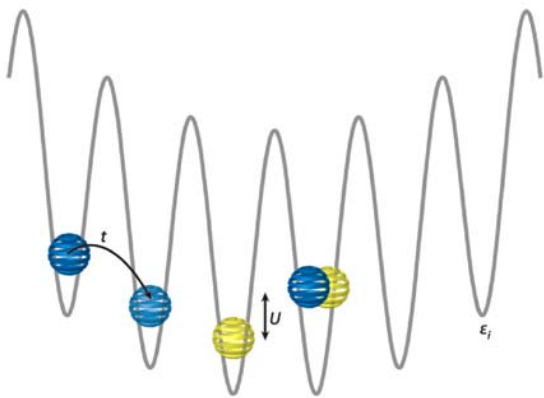

*Figure 2*: A Hubbard model with ultracold atoms. Fermionic atoms trapped in the lowest band of an optical lattice can tunnel between lattice sites with a tunnelling rate *t*. Due to Pauli's principle, tunneling is only possible if the final lattice site is empty or occupied with an atom of a different spin. Two atoms with opposite spin localized at the same lattice site have an interaction energy $U$, which can be likewise positive (repulsive interaction) or negative (attractive interaction). The interplay between $t$ and $U$, the filling and the underlying confining potential determines the physics of the system.

For a two-component gas of fermionic atoms the Hubbard Hamiltonian in an optical lattice reads:

$$H = -t \sum_{\langle i,j \rangle, \sigma} \left( \hat{c}^\dagger_{i,\sigma} \hat{c}_{j,\sigma} + \text{h.c.} \right) + U \sum_i \hat{n}_{i,\uparrow} \hat{n}_{i,\downarrow} + \sum_i \varepsilon_i \hat{n}_i$$

The first term contains the kinetic energy and is proportional to the tunnelling matrix element $t$ between adjacent lattice sites, and <i,j> denotes neighboring pairs. The operators $\hat{c}^\dagger_{i,\sigma}$ and $\hat{c}_{i,\sigma}$ are the fermionic creation and annihilation operators for a particle in the spin state $\sigma$ (up or down) at lattice sites $i$ and $j$. The occupation number of the site $i$ is given by $\hat{n}_{i,\sigma}$. The second term describes the interaction energy in the system and is determined by the on-site interaction $U$. The last term takes account of the additional confinement $V_{\text{trap}}$ of the atom trap, which is usually harmonic. The corresponding energy offset of the lattice site with index $i$ is given by $\varepsilon_i$.

Experimentally, the tunnel coupling is controlled by the intensity of the standing laser waves. This allows for a variation of the dimensionality of the system and it enables tuning of the kinetic energy. A reduced dimensionality is achieved by freezing the atomic motion in certain directions. For example, operating two standing waves at high beam intensities suppresses tunneling in two directions and creates an array of one-dimensional tubes. The energy width of the lowest band is given by $W=4t\mathcal{D}$, where $\mathcal{D}$ is the dimensionality of the system (67).

The on-site interaction $U$ can be approximated by $U = g \int d^3 r \left| w(\mathbf{r}) \right|^4$, where $w(\mathbf{r})$ is the Wannier function of an atom on a single lattice site. Due to the low kinetic energy of the atoms, two atoms of different spin usually interact via s-wave scattering and the coupling constant is given by $g = 4\pi a / m$, with the free space s-wave scattering length $a$ and the atomic mass $m$. The scattering length $a$, and with it the Hubbard $U$, can be tuned to negative or positive values by exploiting Feshbach resonances. An external magnetic field is present in most experiments, which results in a large energy gap between different spin states. In this case, the number of atoms in each spin state is a conserved quantity. A single component Fermi gas is effectively noninteracting because Pauli's principle does not allow s-wave collisions, which are of even parity.

The harmonic confinement provides a trapping potential for the atoms and it is a central aspect of the experimental systems. It is either produced by the optical lattice beams themselves or by an additional optical or magnetic trap. An important consequence of the confinement is that different many-body phases can coexist within the trap. Assuming a local density approximation, the local chemical potential $\mu_{\text{local}}(\mathbf{r})=\mu-V_{\text{trap}}(\mathbf{r})$ decreases with increasing distance $r$ from the trap center, with $\mu$ being the global chemical potential.

For noninteracting fermions, the situation can be analyzed in terms of single-particle energy eigenstates. Low-energy eigenstates, extend around the trap center, and high-energy eigenstates, which are localized in the outer regions of the trap, can be identified (68; 69; 70; 71; 72). A very convenient measure in a harmonic trap with a geometric mean trapping frequency $\bar{\omega}$ is the characteristic atom number $N_0 = \left(W/m\bar{\omega}^2 d^2\right)^{3/2}$ (70; 71). It relates the bandwidth to the trapping potential and defines the number of trapped atoms per spin state, which corresponds to half-filling in the trap center at zero temperature. The related characteristic filling $\rho = N/N_0$ can be controlled in the experiment by changing the total atom number $N$, the trapping frequencies, or the bandwidth.

**Many-Body Physics with Attractive and Repulsive Interactions**

The many-body physics of the Fermi-Hubbard model is governed by the interplay between interaction, delocalization and spin ordering and it covers a wide range of phenomena. In the following I will concentrate on those aspects which are most relevant to the current and near future experimental situations. In optical lattice experiments, it is possible to make all relevant energy scales small compared with the energy gap between the lowest and the next higher Bloch band. The physics is then determined by the energy scales of the bandwidth $W$ and the interaction energy $U$. The next lower energy scale is the super-exchange energy $t^2/U$, which describes virtual tunneling processes. The minimum experimentally achievable temperatures in optical lattices are, at this time, still too high to access this physics of the spin-sector. Due to the trapping potential, there is a further low energy scale, which is the energy separation between adjacent lattice sites.

The attractive-$U$ Hubbard model, where the on-site interaction energy between particles of different spin is negative, has been extensively studied in the context of superconductivity (73). For the homogenous case, i.e., without trapping potential, the general situation is the following. At low temperatures, s-wave superfluidity with a BEC-BCS crossover is expected (74), see figure 3. The BCS regime is characterized by weak attractive interactions between the particles ($U<<t$) and the critical temperature increases with $|U|/t$. In the BEC regime, the strong interactions lead to bound pairs that can undergo Bose-Einstein condensation, with the critical temperature decreasing as $t^2/|U|$. The pairs can be regarded as hard-core bosons, and the tunneling of the pairs is dominated by second order tunneling $t^2/U$. The nearest neighbor interaction between the pairs is repulsive and also proportional to $t^2/U$. For the case of half filling, a charge density wave forms, with a reduced density on every second lattice-site. This leads to a checkerboard pattern in the density distribution (75; 76; 77). Above the critical temperature to superfluidity, a pseudo gap regime of preformed pairs is predicted.

The trapping potential has important consequences for the formation of a charge density wave. The confinement counteracts the nearest neighbor repulsion between attractively bound pairs and favors the formation of a band insulator in the trap center. In order to obtain a sizeable superfluid region inside a trap, a very weakly confining potential and a small number of atoms would be required, i.e., the global chemical potential must not exceed $t^2/U$ (77).

An intuitive starting point for the physics of the repulsive Hubbard model is the situation at half-filling, which maps to the attractive case through particle-hole transformation (17; 76; 77), see figure 3. At higher temperatures, when there is no ordering in the spin degree of freedom, the system exhibits a crossover from a metallic Fermi liquid to a Mott-insulating phase. In the metallic phase, the atoms delocalize to minimize their kinetic energy. In the Mott insulating phase ($U \gg t$), the repulsive interactions between atoms of different spin dominate and the atoms localize. The occupation of a lattice site by two atoms is suppressed, due to either repulsive interactions or the Pauli-principle. The Mott-insulating phase behaves in many ways like a band-insulator and is incompressible. However, the Mott insulator can carry large spin entropy.

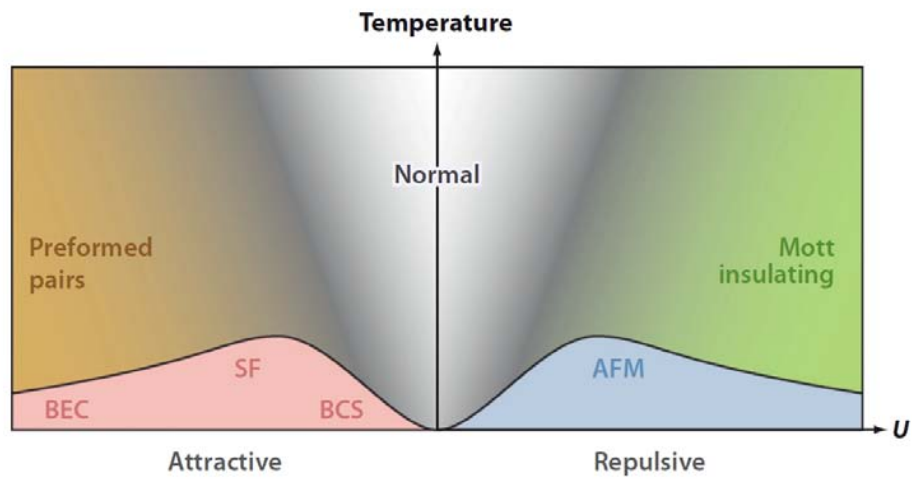

*Figure 3*: Schematic phase diagram for the attractive and repulsive Hubbard model at half filling, for a simple cubic lattice in three dimensions (see, for example Ref. (73)). Abbreviations: BEC, Bose-Einstein condensation; BCS, Bardeen-Cooper-Schrieffer; SF, superfluid; AFM, anti-ferromagnetic phase.

The low temperature phase of the repulsive Hubbard model is characterized by spin ordering. At the Néel temperature, an infinite system undergoes a second order phase transition to the antiferromagnetically ordered state. For weak repulsive interactions, a spin-density wave forms and the Néel temperature is exponentially small in $U/t$ (78; 79). For increasing repulsive interactions, the system exhibits a crossover to an antiferromagnetic Mott insulator. In the Mott-insulating regime, the Néel temperature is proportional to the superexchange energy, which is the dominating scale in the strong coupling limit where $U \gg t$. This can be pictured by consider two adjacent lattice sites, each singly occupied by atoms with opposite spin. Virtual hoping between the sites allows them to lower their energy, compared with the situation in which both atoms have the same spin. This antiferromagnetic superexchange interaction is given by $J_{ex}=4\,t^2/U$ and favors anti-ferromagnetic order. Accordingly, the physics of localized atoms can be described by a Heisenberg spin Hamiltonian $H = J_{ex}\sum_{\langle i,j\rangle} S_i S_j$, where $S_i$ is the spin 1/2 operator (80; 79). The antiferromagnetic order is expected when the system is not magnetically frustrated, which is the case in the typical experimental situation of a simple-cubic lattice. However, it is experimentally possible to create optical lattices of triangular geometry that would frustrate the magnetic ordering. In this case, at very low temperatures, a first-order phase transition between a metal and Mott-insulator is expected (81).

For a qualitative discussion of the local phases inside the trap, we consider the case of strong repulsive interactions with $U \gg T$ and $U \gg W$, while disregarding the superexchange interactions, see figure 4. For a low atom number and a center filling of less than one atom per site, the whole sample is in a metallic phase. An increase in atom number to a level where the global chemical potential exceeds the bandwidth will lead to the formation of a Mott-insulating phase in the trap center. Further increasing the atom number leads to a spatial expansion of the Mott-insulating region until the global chemical potential exceeds the repulsive energy $U$, at which point a metallic phase with a filling larger than one atom per site forms in the trap center. At even higher atom numbers, a band insulator forms in the trap center. It is clear from the previous discussion, that due to the harmonic confinement, it is often not useful to characterize the system by its global properties; therefore one should seek probes sensitive to local properties.

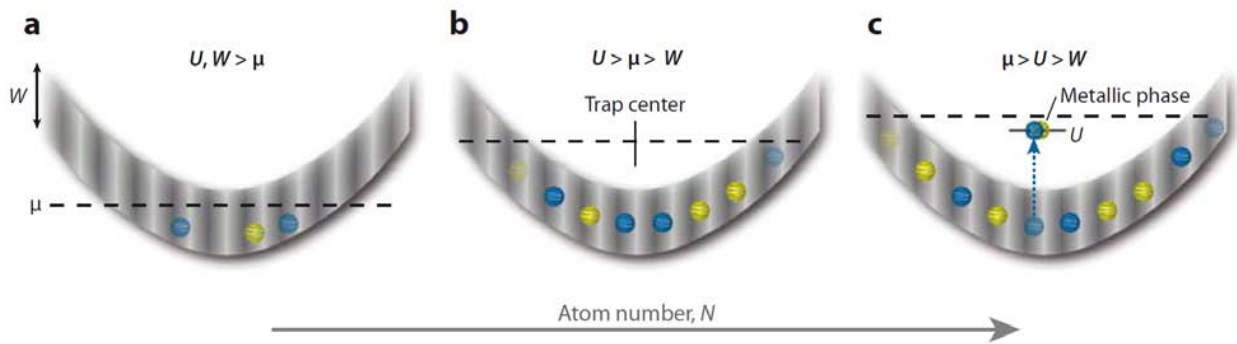

*Figure 4:* The repulsive Hubbard model in a trap with increasing atom number. (a) With less than one atom per site, the whole sample is in a metallic phase. (b) The chemical potential ($\mu$) exceeds the bandwidth and a Mott insulator forms in the trap center. (c) The chemical potential exceeds the local interaction energy $U$ and a metallic phase forms in the trap center. Abbreviations: N, atom number; W, bandwidth.

The excitation spectrum of the repulsive Fermi-Hubbard model changes its character within the cross-over from a metal to an insulator. For small repulsive interactions ($U < t$), the system can be treated as a Fermi liquid and the excitations are wave-like quasi-particles. In the vicinity of the Mott insulator, a gap develops in the energy spectrum and the system can be described in terms of two Hubbard bands. These lower and upper Hubbard bands are energetically separated by the interaction energy $U$ and broadened by hoping processes (82). These high-energy excitations correspond to the creation of doubly occupied lattice sites (doublons). For the ideal case of a magnetically frustrated Mott insulator at zero temperature, it is expected that the spectral weight of the low-energy quasi-particle excitation vanishes at the transition to the Mott insulator and is shifted towards the high energy excitations (17). The high energy excitations are experimentally accessible, as discussed below.

**Limitations of the Fermi-Hubbard Description**

There are several important aspects in the physics of fermionic atoms in optical lattices that are not directly captured by the Fermi-Hubbard model but must be considered. The single-band Hubbard model no longer captures two-particle physics correctly when the scattering length approaches the size of the Wannier function. In this case, higher bands must also be considered and the scattering potential must be properly regularized. For a sufficiently deep optical lattice, individual lattice sites can be approximated by a harmonic oscillator for which the two-particle problem with a regularized contact potential can be solved analytically (83).

Whereas the tunnel coupling determines the shortest time scale in the system, the maximum time scale for an experiment is usually given by atom loss through inelastic collisions between atoms, collisions with background gas atoms, or inelastic scattering of photons from the lattice beams. These processes are not described by the Hubbard Hamiltonian, and they depend on the specific choice of the optical lattice, vacuum conditions and atom species. The lifetime of an atom in an optical lattice is typically on the order of 100 ms to 5 s.

An important consequence of the fact that the trapped atoms are isolated from the environment is that it is most useful to think in terms of entropy per particle instead of in terms of temperature. Because there are presently no cooling methods acting on the atoms inside the optical lattice, the final entropy of the system is provided by the entropy of the ensemble before loading into the optical lattice potential. Assuming adiabatic loading, the total entropy remains the same, but may be reshuffled between different regions of the trap. However, at present, it is not obvious to what degree an initial low-entropy state can be transferred into the optical lattice.

**EXPERIMENTAL REALIZATIONS**

Experiments with fermionic atoms in optical lattices have so far been carried out with the fermionic isotopes of potassium ($^{40}$K) and lithium ($^{6}$Li), as well as fermionic ytterbium ($^{173}$Yb). The Hubbard regime has been reached for $^{40}$K. Lithium experiments have been carried out in a superfluid regime of strong attractive interactions outside the single band description (84), but experiments in the Hubbard regime are under way. In the following, I will briefly discuss the specific properties of $^{40}$K and $^{6}$Li with respect to Feshbach resonances, and I introduce the relevant experimental procedures to produce and observe fermionic quantum gases in optical lattices.

**Feshbach Resonances**

Feshbach resonances, which allow tuning the scattering length in the s-wave channel, are experimentally accessible, both for $^{40}$K and $^{6}$Li. The Feshbach resonance is the result of the coupling between two different interatomic potentials, or channels (39). Resonant behaviour occurs when a bound state of a closed channel is tuned to the energy of the two atoms colliding in the open channel. Control of these energy levels by an external magnetic field is possible if the magnetic moments between the two channels differ. The outcome is a resonant behavior of the scattering length $a$ as a function of the magnetic field $B$. Starting from the background value $a_{bg}$, the scattering length $a(B)$ diverges at the resonance position $B_0$. The scattering length near a Feshbach resonance varies according to $a(B) = a_{bg}\left(1 - \Delta/(B - B_0)\right)$, where the divergence is characterized by its width $\Delta$.

For so-called broad Feshbach resonances, the fraction of the state in the closed channel of the deeply bound molecular state is negligible, and a single-channel description is sufficient to describe the collisional physics. By adiabatically sweeping the magnetic field across the resonance, weakly bound Feshbach molecules of bosonic character can be formed, when starting on the attractive branch. Indeed, using quantum degenerate two-component Fermi gases in harmonic traps, the regime of the BEC-BCS crossover could be studied near the Feshbach resonance (29). The BEC-BCS crossover in the Hubbard regime of an optical lattice is centered at a value of the free attractive scattering length where bound pairs form within the lattice potential.

Several Feshbach resonances are known for $^{40}$K (85; 86; 87). The experimentally most relevant Feshbach resonances are the broad resonances occurring between the magnetic Zeeman sublevels |$F$=9/2, $m_F$=-9/2⟩ and |$F$=9/2, $m_F$=-7/2⟩, located at 202.1 Gauss (6) and between the magnetic Zeeman sublevels |$F$=9/2, $m_F$=-9/2⟩ and |$F$=9/2, $m_F$=-5/2⟩, located at 224.2 Gauss (87). Here $F$ is the total angular momentum and $m_F$ the magnetic quantum number. The background scattering length is positive and has a value of 174 $a_0$ ($a_0$: Bohr radius). The corresponding widths of the resonances are several Gauss. The main limitations for experiments with $^{40}$K are the inelastic losses encountered in the vicinity of both Feshbach resonances. For an attractively interacting potassium gas in the |F=9/2, $m_F$=-9/2> and |F=9/2, $m_F$=-7/2> mixture this limits the lifetime of the superfluid phase in the BEC-BCS crossover in a harmonic trap to around 30 ms (6), even though the pair lifetime is enhanced by Pauli blocking (88). The lowest reported temperature in a harmonic trap for an attractively interacting gas of $^{40}$K is $T/T_F$=0.07 (6). The positive background scattering length makes it straightforward to efficiently cool the gas at strong attractive interactions and then to tune the scattering length to the repulsive background value.

The experimentally most important Feshbach resonance in $^6$Li occurs between the two lowest hyperfine states, |F=1/2, $m_F$=1/2> and |F=1/2, $m_F$=-1/2>, and is located at 834.15 Gauss. It has a width of 300 Gauss and the background scattering length is negative with $a_{bg}$=-1405 $a_0$ (89). Feshbach molecules formed by sweeping over the Feshbach resonance have a remarkably long lifetime. The long lifetime is attributed to the exceedingly small fraction in the closed channel. This has enabled a range of experiments with a lithium gas in the unitary regime. The lowest reported temperature for an attractively interacting gas of $^6$Li is $T/T_F$=0.05 (30).

**Preparing the Quantum Gas in the Optical Lattice**

Before loading the quantum gas into the optical lattice, it is evaporatively cooled in a magnetic trap, or an optical dipole trap, and prepared in the desired spin states. Then the gas is transferred into the optical lattice by ramping up the lattice potential within 50 ms to 200 ms. This time scale is chosen to be long enough to avoid excitations into higher bands and short enough to avoid substantial inelastic losses and heating due to spontaneous emission. It is often assumed that the many-body wave function of the quantum gas can follow the changing potential adiabatically. Experimentally, it has be shown that a sample at an initial temperature in the dipole trap of $T/T_F$=0.15 can be loaded into the optical lattice and transferred back into the dipole trap resulting in a final temperature of $T/T_F$=0.2 - 0.24, with lower spreads obtained for weaker interactions (23; 24). Yet there are no systematic studies in different regimes of parameters, such as interaction strength and characteristic filling or trap geometry. Methods to determine the temperature of the gas inside the optical lattice are currently being developed or evaluated (90; 91; 92).

**Time of Flight Expansion Imaging**

A central method to obtain information from the atomic sample is absorption imaging after a period of ballistic expansion (32). In this method, the optical lattice and the harmonic trapping potential are switched off, and after a ballistic expansion period of typically 5 to 30 ms, the atoms are illuminated by a probe laser that is resonant with the atomic transition. The shadow cast by the atoms is then imaged onto a CCD camera. From the resulting image, the optical depth of the expanded cloud integrated along the imaging axis is obtained and the total atom number can be extracted. The calibration of an absolute number of atoms is typically limited to an accuracy of ±10% (93). For the case of a sudden switch off, i.e., much shorter than the inverse on-site trapping frequency, the atomic distribution can be interpreted as an image of the momentum distribution of the atoms inside the optical lattice. The sudden switch off leads to a projection of the quasi-momentum states in the lattice onto real momentum states in free space. Correspondingly, the images have the structure of an interference pattern. The effective momentum resolution is determined by the initial cloud size and the duration of the time of flight expansion. A related method is the adiabatic switch off (94). Here the optical lattice potential is gradually lowered, such that the quasi-momentum distribution inside the optical lattice is mapped on the real momentum states in free space. Both methods are affected by the inhomogenity of the trapping potential due to averaging of the momentum distribution over the whole cloud and interaction during the early stage of the expansion.

A very powerful addition to time-of-flight imaging is to expose the atoms during ballistic expansion to an inhomogenous magnetic field. Through the Stern-Gerlach effect, this results in a force acting on the atoms that is sensitive to the magnetic moment of the atomic state. Correspondingly, atoms with different magnetic moments separate during ballistic expansion (85). This allows one to independently measure the momentum distribution or the total number of atoms for each spin component.

An alternative to ballistic expansion is the imaging of the atomic cloud inside the trap. Due to the high atomic densities, the clouds are often optically thick, making absorption imaging difficult. This can be circumvented by applying phase contrast imaging (32). This method is very effective when the imaging system can resolve spatial structures much smaller than the atomic cloud.

## EXPERIMENTS WITH FERMIONIC QUANTUM GASES IN THREE-DIMENSIONAL OPTICAL LATTICES

### Non-Interacting Fermi-Hubbard System

In the first experiment implementing a Fermi-Hubbard model, a $^{40}$K gas was loaded into a three-dimensional optical lattice with a wavelength of 826 nm (19). A two-component mixtures was realized by equally populating the spin states $|F=9/2, m_F=-9/2\rangle$ and $|F=9/2, m_F=-7/2\rangle$ in the gas. After loading the gas into the lowest band of an optical lattice, a homogenous magnetic field of 210 Gauss was applied to tune the scattering length between the two spin components from the background value to its zero crossing. Using this method, a noninteracting Fermi degenerate quantum gas inside the optical lattice was realized. Direct images of the Fermi-surfaces were possible by slowly ramping down the optical lattice potential followed by ballistic expansion and absorption imaging, see figure 5. In the experiments, the evolution from a metallic sample to a band insulator could be followed by increasing the characteristic filling. Furthermore, turning on strong repulsive interactions between the two spin states resulted in a coupling between the two lowest bands. The observed shape of the Fermi surfaces can be regarded as a macroscopic result of the Pauli-principle, yet the inhomogeneity of the system and finite temperature prevents a direct measurement of the Fermi-edge.

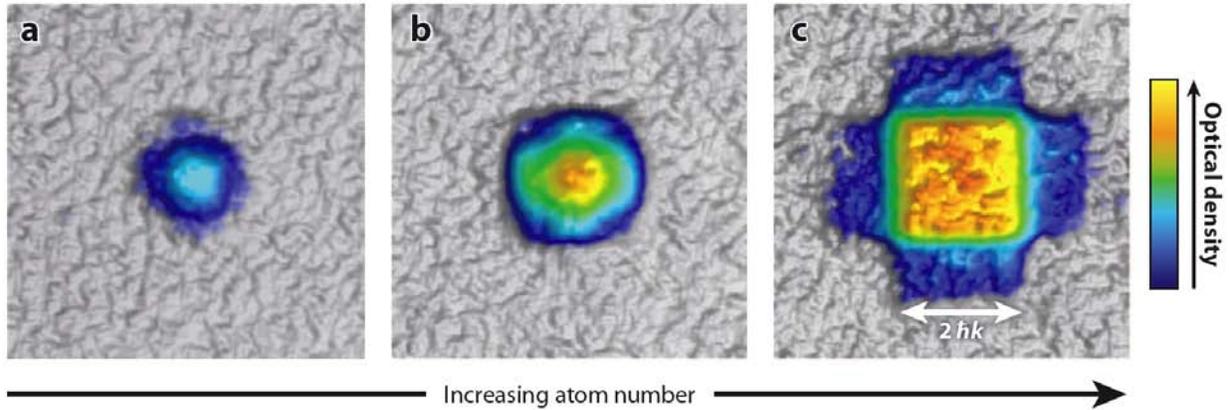

*Figure 5*: Observing the Fermi surface. The absorption images are taken after adiabatically lowering the lattice potential followed by 9 ms of ballistic expansion. The false color distributions show the optical density of the expanded atomic cloud along the imaging direction. They reflect the quasimomentum distributions of the atoms in the optical lattice projected on the imaging plane. The boundary of the distribution reveals the Fermi surface. The initially trapped atom number increases from left to right. (a) For a low atom number, the Fermi surface is of spherical shape and the corresponding projection is seen. (b) With increasing atom number, the Fermi surface develops a more complex structure, showing extensions toward the boundary of the first Brillouin zone. (c) The filling of the lattice in the trap center exceeds unity such that higher bands are populated, which show up beyond the square shape of the first Brillouin zone (see also Reference (19)).

Microscopic access to the fermionic nature of a spin polarized $^{40}$K quantum gas loaded into an optical lattice has been achieved by analyzing the correlations in the noise of absorption images, which were taken after a sudden lattice switch off and a period of ballistic expansion (95; 96). The sudden switch-off of the optical lattice acts like a beam splitter and projects a Bloch state with quasi-momentum $q$ onto its momentum components $q+n\hbar k$ in the free space basis (n=0, ±1,...). Due to the Pauli principle, only a single atom can populate each Bloch state, and its detection in one free space momentum component will exclude its detection in the other momentum components (97). Absorption imaging integrates the atomic distribution along one axis and substantially reduces this anti-correlation, but it could clearly be detected in the experiment with correlation amplitudes of $-4\times10^{-4}$.

**Feshbach Molecules and Lattice Bound Pairs**

One can distinguish between two forms of bound two-particle states in an optical lattice. There are attractively and repulsively bound states that are stabilized by the optical lattice potential. These localized states exist due to the presence of the lattice potential, and their energies are above or below the continuum of the lowest Bloch band (98). In addition, there are the Feshbach molecules, which also exist in free space without the optical lattice. Using a two-component $^{40}$K gas in a three dimensional optical lattice, the transition from attractively bound pairs to Feshbach molecules has been investigated experimentally (90). The quantum degenerate gas was prepared with a large number of lattice sites being occupied by two atoms, one in each of the two spin components |$F$=9/2, $m_F$=-9/2> and |$F$=9/2, $m_F$=-7/2>. Starting near the zero crossing for the scattering length, the Feshbach resonance was used to increase the attractive interaction between the atoms and to create Feshbach molecules. The binding energy of both types of pairs was spectroscopically measured using a radio frequency transition into a third spin state (|$F$=9/2, $m_F$=-5/2>). Good quantitative agreement with a localized harmonic oscillator model was obtained (83). It was also shown that the attractively bound pairs dissociate when the optical lattice is switched off, whereas the Feshbach molecules remained stable. A complementary study for repulsively interacting particles has been carried out for bosonic atoms in an optical lattice, where repulsively bound pairs could be identified (98).

Further insight into the physics of local pairs has been gained by probing the fraction $D$ of atoms residing on doubly occupied lattice sites for the noninteracting case and for various strengths of attractive interactions. For the noninteracting gas, the double occupancy in the lattice is solely determined by the number of trapped atoms - corresponding to a certain characteristic filling - and the temperature of the sample. This method can therefore be used as a thermometer for the non-interacting sample (22; 91). An increase in double occupancy is observed when turning on and gradually increasing the attractive interactions between the atoms in the optical lattice. Saturation of double occupancy at a maximum value of 60% was measured and found to set in for attractive interactions beyond $U/t \approx -8$, at a value when attractively bound pairs start to form (73).

**Superfluid of Ultracold Fermions in an Optical Lattice**

The coherence properties of a fermionic superfluid in an optical lattice were studied with a quantum degenerate gas of $^6$Li atoms (84). Using the Feshbach resonance at 834 G, the two-component gas was prepared in the regime of the BEC-BCS crossover near-unitary scattering and thus outside the Hubbard regime. The experiment addressed the question to which extent the condensed pairs would retain their coherent nature when exposed to the optical lattice potential. The periodic potential was formed by three standing laser waves at 1064 nm and was gradually imposed on the gas. To probe the momentum distribution of the pairs in the lattice, the magnetic field was rapidly ramped out of the strongly interacting region and the lattice potential was switched off. After ballistic expansion, absorption images were taken, which showed the distribution of the Feshbach molecules formed during the magnetic field ramp, see figure 6. The level of coherence of the sample inside the lattice was deduced from the width of the interference peaks observed in the images of the momentum distribution. It was found that the superfluid could retain its long-range coherence up to a critical lattice depth. For the fermionic atoms, this critical lattice depth corresponded to $V_0 = 3\ E_r$, or twice this value for the Feshbach molecules. Due to this comparatively low lattice depth, the strong interactions, and a central density of one pair per lattice site, a single-band Hubbard model does not capture the observed physics. In fact, a gap in the single-particle band structure of the lattice opens up only when $V_0 / E_r$ > 2.23 (99). The observations can be described in terms of multiband models displaying an underlying quantum phase transition from a superfluid to an insulating state (99; 100; 101; 102). For attractive interactions, in the BCS limit the insulating state can be interpreted as a band insulator of fermions, and in the BEC regime, it can be interpreted as a Mott insulator of molecules.

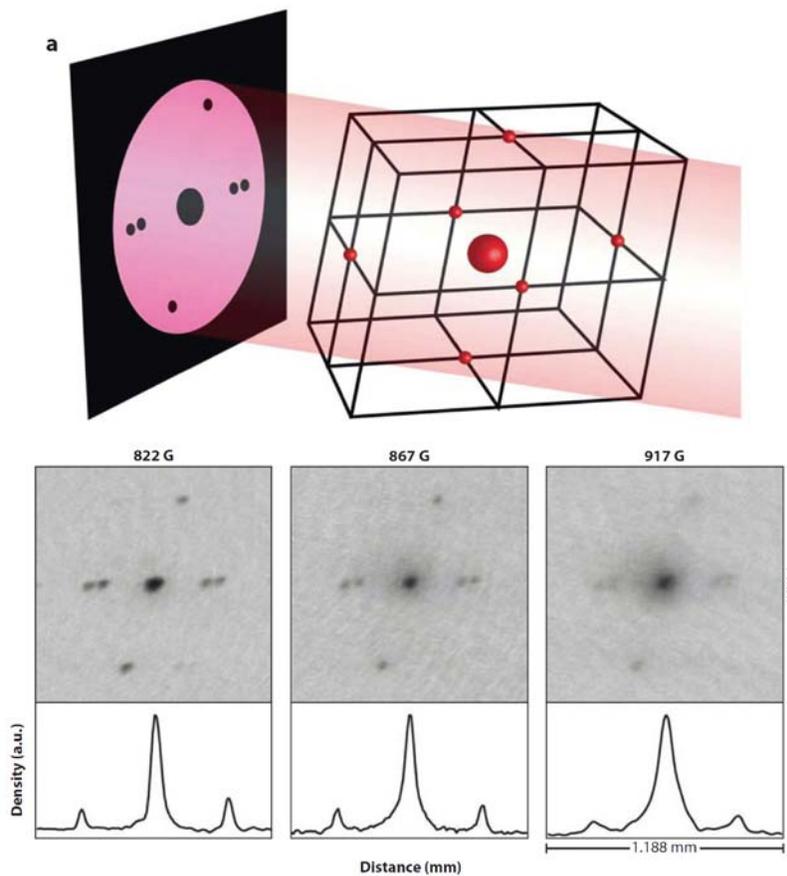

*Figure 6*: Observation of high-contrast interference of fermionic pairs of $^6$Li atoms released from an optical lattice below and above the Feshbach resonance. (a) The orientation of the reciprocal lattice, also with respect to the imaging light. Interference peaks are observed for magnetic fields of 822 G (b), 867 G (c) and 917 G (d). The lattice depth for pairs is 5 $E_r$ in all images, and each image is the average of three shots. The field of view is 1 mm x 1 mm. Density profiles through the vertical interference peaks are shown for each image. The horizontal axis for the density profiles is a tilted axis through the interference pattern; thus, the measurement of the frame size has been adjusted for axis tilt. Figure adapted from Reference (84).

**The Mott Insulator and Repulsive Interactions**

In solid materials, the Mott insulating phase leads to a suppression of conductivity, which in this case is a result of interactions rather than of a filled Bloch band (103). The proximity to the Mott insulating phase is the origin of many intriguing phenomena in condensed matter physics. In atomic systems, the physics of a Mott insulator was first encountered in the observation of the superfluid to Mott-insulator transition when a Bose-Einstein condensate was loaded into a three-dimensional optical lattice of variable depth (12). In the superfluid phase, the long-range phase coherence results in narrow peaks in the momentum distribution, and their disappearance indicates the transition to the Mott insulator. This method cannot be usefully applied to a fermionic quantum gas in an optical lattice, since the kinetic energy changes only marginally during the cross-over from a metallic to a Mott insulating phase (70).

Instead of measuring the kinetic energy in the fermionic system, a change in interaction energy can be sensitively probed by measuring the fraction of doubly occupied lattice sites $D$ (23). This method is highly responsive to the trap center were the double occupancy first starts to build up. It is also related to the core compressibility (104). In the experiment, a two-component $^{40}$K Fermi gas is loaded from an optical dipole trap into a three-dimensional optical lattice operated at $\lambda$=1064 nm. Using the Feshbach resonances at 202.1 Gauss and 224.21 Gauss, the initial many-body state could be prepared over a wide range of interactions, from vanishing ($U/6t$ = 0) to strong repulsive interactions ($U/6t$ = 30) and at lattice depths between 7 $E_r$ and 12 $E_r$. To probe the prepared state, the atoms were abruptly frozen to their lattice sites by suppression of tunneling through a sudden increase of the lattice depth to 30 $E_r$. In the next step, approaching a Feshbach resonance shifted the energy of atoms on doubly occupied sites, so that a radiofrequency pulse could selectively transfer one spin state to a third one. The latter spin state is a previously unpopulated magnetic sublevel. The fraction of transferred atoms and the resulting double occupancy is then obtained from absorption images of all three spin components using Stern-Gerlach separation.

The measured double occupancy as a function of the total atom number is shown in figure 7. For the noninteracting case the anticipated rapid increase in double occupancy with atom number is detected. For strong repulsive interactions, a pronounced suppression of double occupancy is clearly visible until double occupancy eventual builds up for larger atom numbers. For repulsive interactions that are larger than the temperature, this behavior is expected when the atom number, i.e., the chemical potential, is increased: First, the lowest Hubbard band is filled and a Mott insulator forms, which then increases in size until it starts to melt in the trap center and double occupancy builds up. The measured results could be fitted well with calculations of a Hubbard model in the atomic limit, which neglects tunneling. The Mott insulator could be characterized by directly measuring the double occupancy (< 2%), and by deducing the number of holes (< 3%) from a realistic estimate of the temperature. For the largest interaction strength, a ratio of $k_B T/U$=0.11 was reached, with $k_B$ being Boltzmann's constant. More recently, a quantitative measurement of the double occupancy with minimized systematic errors was carried out for interactions strengths between $U/6t$ = 1.4 and $U/6t$ = 4.1 at a lattice depths of 7 $E_r$. Very good quantitative agreement with ab initio calculations was found (105) for dynamical mean field theory (17) and for results from a high-temperature expansion (104).

Measurements of the cloud size have been used as a further tool to investigate the effect of repulsive interactions in an atomic realization of the Fermi-Hubbard model (24). A two-component spin mixture of $^{40}$K atoms was loaded into a three-dimensional optical lattice in which the harmonic confinement could be varied independently from the depth of the optical lattice. This was achieved by operating the optical lattice at a wavelength of 738 nm, where the potassium atoms are trapped in the intensity minima of the standing waves. Thus, the compression of the cloud could be decreased not only by decreasing the atom number but also by decreasing the trapping frequencies of the optical dipole trap operated at 1030 nm. The change in the cloud radius was extracted from phase contrast images that recorded the projection of density distribution of the trapped gas.

Using this method, it was found that the noninteracting gas displays the expected rapid decrease of cloud radius under increasing compression until a band insulator forms in the center of the trap and counteracts further compression. With increasingly repulsive interactions, larger cloud sizes were measured for the same compression. Agreement with theoretical predictions using dynamical mean field theory was found for non-interacting and weakly interacting systems ($U/6t \leq 2$). For intermediate interactions ($U/6t$ = 3), still larger cloud sizes were measured, and the experimental data were found to be consistent with the formation of an incompressible Mott-insulating core.

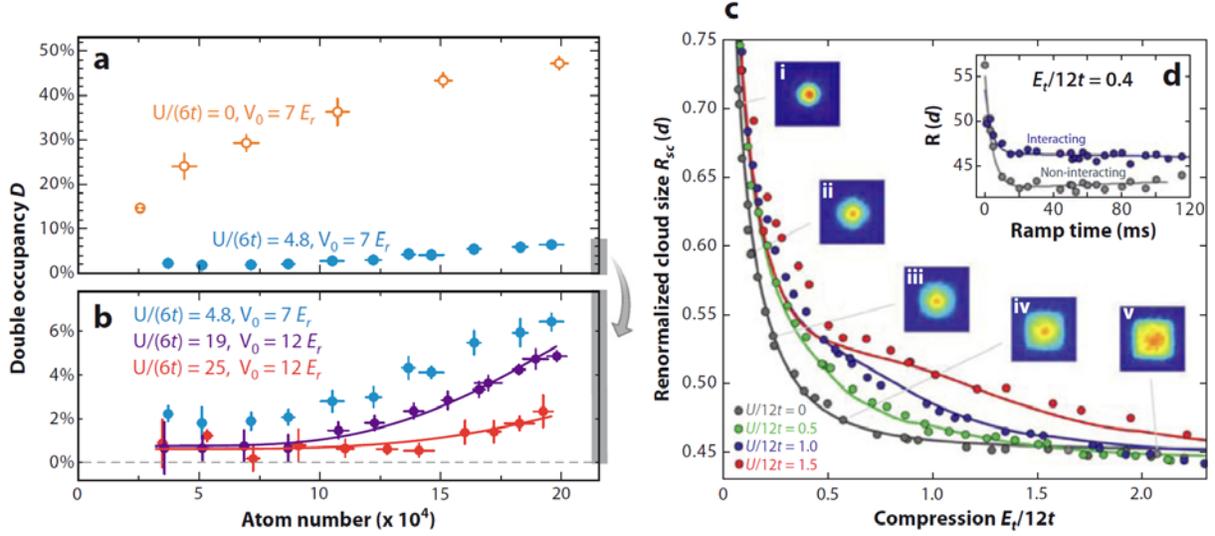

*Figure 7*: (a,b) Double occupancy measured in the non-interacting and Mott insulating regimes. (a) A significant increase in the double occupancy with atom number is observed in the noninteracting regime (open orange circles), whereas repulsive interactions suppress the double occupancy (filled blue circles). (b) In the Mott insulating regime, the double occupancy is strongly suppressed. It starts to increase for large atom numbers, indicating the formation of a metallic region in the trap center. The purple and red lines represent the theoretical expectation for $D$ in the atomic limit. Figure adapted from Reference (23). (c) Cloud sizes of the interacting spin mixture versus compression. Measured cloud size $R_{sc}$ in a $V_{lat} = 8\,E_r$ deep lattice as a function of the external trapping potential for various interactions $U/12t = 0...1.5$ is shown. Dots denote single experimental shots, lines denote the theoretical expectation from dynamical mean-field theory calculations for an initial temperature $T/T_F = 0.15$. The insets (i to v) show the quasi-momentum distribution of the noninteracting clouds (averaged over several shots). (d) Resulting cloud size for different lattice ramp times at $E_t/12t = 0.4$ for a noninteracting and an interacting Fermi gas. The arrow marks the ramp time of 50 ms used in the experiment. Adapted from Reference (24).

**Hubbard Excitations: Spectroscopy and Lifetimes**

The energy separation between the two Hubbard bands has been measured (23) using modulation spectroscopy (62; 106). In this method, a periodic modulation of the lattice intensity at a frequency $\omega$ induces extra double occupancy, which is then detected. A strong increase in double occupancy is observed near the resonance condition $\omega = U/\hbar$, see figure 8. The mechanism can be understood in perturbation theory (106; 107; 108). The modulation of the optical lattice potential $[V(\tau)=V_0+\delta V\sin\omega\tau]$ results in a periodic modulation of the tunneling matrix element $t$ and on-site interaction $U$, with the modulation of the tunneling term being the relevant contribution. Up to quadratic response, the absorbed energy results in the creation of double occupancy. This response occurs at an energy $U$ and there is no response at the low-energy scale of metallic excitations.

A general characteristic of an atomic quantum gas is that it is not coupled to the environment or a thermal bath. Relaxation of an excited state therefore occurs through internal redistribution of energy. Due to the large energy gap in the excitation spectrum, the doublon excitations are strongly protected against decay. Recently, the lifetime of doublon excitations has been measured experimentally (109). It was found that their lifetime scales exponentially with $U/6t$ over a large parameter range. This indicates that high-order scattering events are the dominant decay process. In these multiparticle scattering processes, each scattering event can contribute at most the energy of the bandwidth. Fair agreement with the experimental data was obtained in a diagrammatic approach.

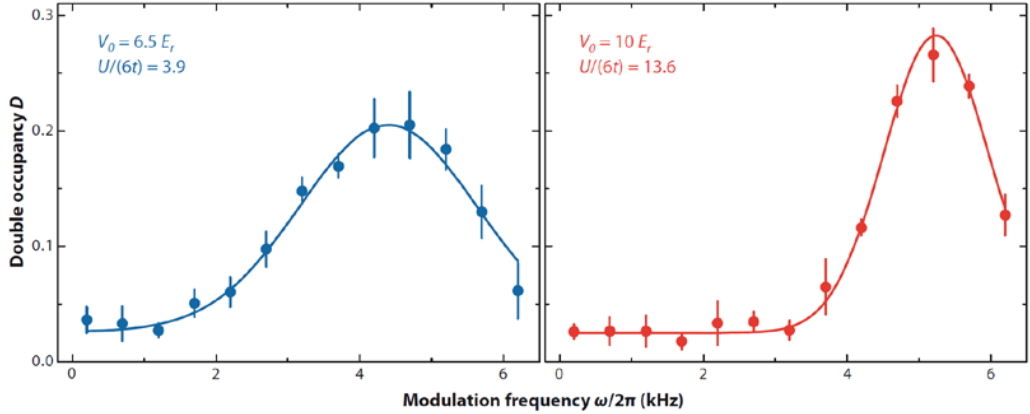

*Figure 8*: Modulation spectroscopy. In the regime of strong repulsive interactions, a periodic modulation of the lattice intensity at a frequency $\omega$ induces additional double occupancy $D$. The resonance occurs at the energy separation between the two Hubbard bands, i.e., $\omega = U/\hbar$. Figure adapted from Reference (23).

**TOWARDS ANTI-FERROMAGNETISM AND BEYOND**

The next major step will be to reach the regime of anti-ferromagnetic ordering, which is an important milestone towards a d-wave superfluid phase. Furthermore, this regime would give experimental access to intriguing questions of quantum magnetism. At present, it is not obvious whether spin ordering is attainable with current methods and existing experimental set-ups. From a recent quantitative measurement in the relevant regime, it was concluded that the entropy per particle is a factor 2 above the expected onset of an anti-ferromagnetic phase in the Heisenberg model (105). The challenge is the following: Temperatures that correspond to the minimally attainable entropies per particle are reached for large attractive interactions between the particles and not for repulsive interactions, as required for antiferromagnetic ordering. Strong repulsive interactions in the vicinity of a Feshbach resonance are accompanied by inelastic losses. This gives rise to heating and limits the time scale on which experiments can be performed. In addition it is not clear how close one can stay to a fully adiabatic path when turning on the optical lattice potential.

These temperature limitations of present experiments have been addressed in several proposals that suggest new routes to lower temperatures in optical lattices. A pragmatic approach is to exploit the low temperatures for attractive interactions and to study the physics of the attractive Hubbard model, which maps to the repulsive models in a variety of regimes (76; 77). The removal of entropy from the sample is a further route for future experiments. Methods have been suggested to selectively address and remove defects in an optical lattice (110). It has also been realized that the entropy is unevenly distributed within the trap (111; 112), for example, a band insulating region in the center of a trap is free of defects (68) and cannot take on entropy. Cooling schemes have therefore been proposed in which localized regions of low entropy are isolated from the rest of the system (111; 113; 114) . The creation of low entropy regions may also be possible by suitably shaping the confining potential (115; 116).

Another approach is to bring the lattice gas into thermal contact with a second species that may then cool the sample (117; 118). Such entropy exchange has been suggested to be responsible for an observed reduction in coherence of a superfluid in an optical lattice when mixed with fermions (20), and this was directly studied in a recent experiment (119). Using ideas from dark state laser cooling, a scheme was recently proposed that would allow it to cool fermionic atoms immersed in a superfluid to temperatures below the reservoir temperature (120). An application of an initial low-entropy state is to adiabatically transform a well-defined initial state on a plaquette into a d-wave resonating valence bond state (121).

Reaching lower temperatures and entropies is a fundamental frontier. The development of novel probes for the systems is more technical in nature but of similar importance. From a physics point of view, the challenge is to extract relevant correlation functions from the system. Ideally, one would like to be able to extract critical exponents from measurements in the vicinity of a phase transition (122). Again, different approaches have been proposed and some of them are presently being tested. A very effective extension of time-of-flight imaging is the analysis of the correlations in noise of these images (96; 95; 123; 124), which should be sensitive to anti-ferromagnetic ordering. There are also various ideas that aim to exploit the scattering of light from the sample. A promising possibility to observe the antiferromagnetic ordering is to use Bragg-scattering of light from the lattice sample (125), as has been employed for laser cooled atoms (53; 52). The probing of reduced spin-noise, due to pairing or ordering, measured by passing a focused beam through the sample, has also been suggested (126).  A further step in this direction is the use of cavities (127; 128; 129) or interferometers (130) to enhance the coupling between the many-body sample and the light field.

A different route is the probing of the response of the sample to an external perturbation, such as lattice or trap modulation (62; 106; 131), and Bragg- or Raman spectroscopy (132; 133; 134; 135; 136). These methods become particularly powerful once the regime of linear response is reached.  A technique based on Raman spectroscopy to probe the one-particle Green function and the Fermi surface in the strongly correlated regime has recently been analyzed in detail (137). A related experiment using radio frequency transitions has recently been carried out with a Fermi gas in a harmonic trap (138).

Measurements of density-density correlations at the single atom level have been realized in quantum gas experiments (139; 140; 97), and techniques to probe single atoms on single sites have recently been implemented in optical lattices (141; 142). The power of these techniques has been demonstrated by observing incompressible Mott-insulating domains in situ of an optical lattice (143). Furthermore, high-resolution electron microscopy of atoms in an optical lattice could be demonstrated successfully (144).

Optical lattice potentials are not limited to the presently predominant simple cubic configuration. Potentials of a more complex periodicity, such as superlattices (145; 146) or triangular lattices, can be created and dynamically controlled (147). Almost arbitrary structures can be envisaged for two-dimensional systems by imaging suitable potentials. Furthermore, potentials with controlled disorder have been implemented recently in bosonic quantum gas experiments (148; 149; 150)

There is an enormous range of physics within and beyond the Fermi-Hubbard model which will become accessible in experiments. In my opinion, the major themes will be magnetic ordering and magnetic frustration, doping, and disorder, as well as long-ranged interactions, meta-stable states and non-equilibrium physics.

Acknowledgements: I acknowledge many stimulating discussions with E. Altman, J. Blatter, I. Bloch, H.P. Büchler, M. Cazallila, E. Demler, A. Georges, T. Giamarchi, D. Greif, M. Greiner, K. Günter, T.-L. Ho, S. Huber, R. Jördens, M. Köhl, C. Kollath, H. Moritz, L. Pollet, M. Rigol, M. Sigrist, T. Stöferle, N. Strohmaier, L. Tarruell, P. Törmä, M. Troyer, T. Uehlinger, W. Zwerger, and many others. The work is financially supported by the Swiss National Science Foundation, ETH Zurich and the EU (SCALA and NAMEQUAM ). The project NAMEQUAM acknowledges the financial support of the Future and Emerging Technologies (FET) program within the Seventh Framework Program for Research of the European Commission, under FET-Open grant number: 225187.